# Hidden States and Dynamics of Fractional Fillings in tMoTe$_2$ Moiré Superlattices

Yiping Wang[1,2], Jeongheon Choe[1], Eric Anderson[3], Weijie Li[3], Julian Ingham[4], Eric A. Arsenault[1], Yiliu Li[1], Xiaodong Hu[5], Takashi Taniguchi[6], Kenji Watanabe[7], Xavier Roy[1], Dmitri Basov[4], Di Xiao[5], Raquel Queiroz[4], James C. Hone[2], Xiaodong Xu[3,5], X.-Y. Zhu[1,*]

[1] Department of Chemistry, Columbia University, New York, NY 10027, USA

[2] Department of Mechanical Engineering, Columbia University, New York, NY 10027, USA

[3] Department of Physics, University of Washington, Seattle, WA 98195, USA

[4] Department of Physics, Columbia University, New York, NY 10027, USA

[5] Department of Materials Science and Engineering, University of Washington, Seattle, WA 98195, USA

[6] Research Center for Materials Nanoarchitectonics, National Institute for Materials Science, 1-1 Namiki, Tsukuba 305-0044, Japan

[7] Research Center for Electronic and Optical Materials, National Institute for Materials Science, 1-1 Namiki, Tsukuba 305-0044, Japan

**The fractional quantum anomalous Hall (FQAH) effect was recently discovered in twisted MoTe$_2$ bilayers (tMoTe$_2$)[1–4]. Experiments to date have revealed Chern insulators from hole doping at ν = -1, -2/3, -3/5, and -4/7 (per moiré unit cell) [1–6]. In parallel, theories predict that, between $v$ = -1 and -3, there exist exotic quantum phases [7–15], such as the coveted fractional topological insulators (FTI), fractional quantum spin Hall (FQSH) states, and non-abelian fractional states. Here we employ transient optical spectroscopy [16,17] on tMoTe$_2$ to reveal nearly 20 hidden states at fractional fillings that are absent in static optical sensing or transport measurements. A pump pulse selectively excites charge across the correlated or pseudo gaps, leading to the disordering (melting) of correlated states [18]. A probe pulse detects the subsequent melting and recovery dynamics via exciton and trion sensing [1,3,19–21]. Besides the known states, we observe additional fractional fillings between ν = 0 and -1 and a large number of states on the electron doping side (ν > 0). Most importantly, we observe new states at fractional fillings of the Chern bands at ν = -4/3, -3/2, -5/3, -7/3, -5/2, and -8/3. These states**

---

**are potential candidates for the predicted exotic topological phases** [7–15]**. Moreover, we show that melting of correlated states occurs on two distinct time scales, 2-4 ps and 180-270 ps, attributed to electronic and phonon mechanisms, respectively. We discuss the differing dynamics of the electron and hole doped states from the distinct moiré conduction and valence bands.**

Quantum states that support fractional charge excitations are fascinating manifestations of many-body interactions in condensed matter. These states, originally discovered as the fractional quantum Hall (FQH) effect in 2D electron gases at high magnetic fields[22], have been observed recently in moiré superlattices of tMoTe$_2$ bilayers [1–4] and multilayer graphene/hexagonal boron-nitride interfaces [23] in the absence of magnetic field. Unlike fractional charge orderings from generalized Wigner crystal formation in moiré bands [20,21,24,25], the recently discovered FQAH effects[1–4,23] represent fractional charge excitations obeying anyon statistics, formed from many body correlations in flat Chern bands. In addition to the FQAH states, fractional fillings between $v$ = -1 and -3 have been predicted to result in other exotic quantum phases [7–15], but these states have not been detected to date. A transport measurement assigned a $v$ = -3 state to FQSH state[26], but the presence of ferromagnetic order may challenge this interpretation [27,28].

To find new states in the Chern bands, we investigate dual-gated R-stacked MoTe$_2$ bilayers with twist angles of θ = 3.7º and 3.1º, respectively (Fig. 1a, detailed in Extended Data Fig. 1 and 2), known to exhibit FQAH effects [1–4]. In our approach [16,17], illustrated in Fig. 1b for a Mott insulator, a pump pulse with $\hbar\omega_1$ (= 0.99 eV) below the optical gap (1.12 eV) [1] of MoTe$_2$ induces intraband excitation of a doped electron or hole. Relaxation within the continuum of conduction or valence band states occurs on ultrafast (≤ 100 fs) time scales (II) [16,17,29], but the correlation gap provides a bottleneck in relaxation and leads to the transient formation of excitation across the gap (III), referred to as a holon-doublon pair in a Mott insulator [18]. The presence of sufficient excitation leads to disordering and gap-closing [16–18] (IV), followed by recovery at longer times (V). This approach also applies to correlated electron liquid, where the bottleneck in intraband relaxation leads to the transient formation of excitation across a pseudo-gap, thus disrupting correlation and renders the system more Fermi liquid like. The dynamics are followed as a function of pump-probe delay (Δt) from exciton (~1.120-1.135 eV) and trion (~1.100-1.120 eV) sensing by the probe pulse ($\hbar\omega_2$), which detects the increase in the effective dielectric constant [19–21] and the resulting decrease in oscillator strengths. We present $\Delta R/R$, where $\Delta R = R(\Delta t) - R$; $R(\Delta t)$ and $R$ are the reflectance spectrum at $\Delta t$ and before time zero (static), respectively. The pump-induced change is detected as a reduced amplitude of the derivative-shaped resonance and $\Delta R/R$ shows a characteristic flip in sign from that of R (Fig. 1c). Our measurement reveals predominantly the equilibrium states before they are "melted" by the pump pulse. We do not observe increases in oscillator strength after photo-excitation, indicating transient formation of correlated insulators are not important. The sensitivity and specificity of pump-probe spectroscopy comes from its background-free nature, because the correlated states are selectively perturbed by the pump pulse when the photon energy is below the optical gaps of constituent 2D semiconductors.

**Time domain detection of fractional fillings of the Chern bands: $\theta = 3.7°$**

We first focus on sample D1 ($\theta = 3.7°$) and compare the static reflectance spectrum (Fig. 1d) and transient spectra, Fig. 1e and 1f at $\Delta t$ = 13 ps and 300 ps, respectively, as functions of gate voltage ($V_g$) and $\hbar\omega_2$. In static spectrum, the $\nu = -1$ and 1 states can be identified by the enhancements in R. The transient spectra show drastically enhanced resolution and sensitivity, leading to the detection of 20 states. Fig. 1g shows selected linecuts at exciton (black) and trion (orange) resonances, with the moiré filling factor $\nu$ listed (see Method and Extended Data, Table 1). A static spectroscopic technique based a separate $WSe_2$ sensor layer was used to detect a large number of correlated insulators in the topologically trivial $WS_2/WSe_2$ moiré superlattice[21], but this approach is less sensitive for the $tMoTe_2$ system[3].

On the electron side, we resolve $\nu = 1$ and at least 10 states at fractional fillings. The number of states identified here is more than twice of those detected in steady-state spectroscopic sensing[1]. These states, also observed in other transition metal dichalcogenide (TMD) moiré systems, have been called generalized Wigner crystals [20,21,24,25]. They are formed when the onsite ($U$) and intersite ($V$) Coulomb energies overwhelm the kinetic energy ($t$). In $WS_2/WSe_2$ moiré superlattices, Arsenault et al. recently pointed out the similarity of generalized Wigner crystals to charge density waves (CDW) and showed evidences for the polaronic nature of localized charges[16].

From hole doping of the Chern bands, we observe states at $\nu = -1, -2/3, -1/2, -1/3$, and likely $-3/5$ and $-2/5$. These states have been predicted in the phase diagrams for hole doping[30]. The $\nu = -1, -2/3, -3/5$ states are zero-field Chern insulators [1–4] and $\nu = -1/2$ a composite Fermi-liquid[31]. The

ν = -1/3 state is in the region without magnetic order [4] and is a topologically trivial charge order state[8,30,32]. The ν = −2/5 state is also in the non-magnetic region [28] and possibly a charge order state. For doping beyond ν = -1, we observe new fractional states at ν = -4/3, -3/2, and -5/3. Each of these new states exhibits a small feature on top of intense ΔR/R signal over broad doping levels (ν < -1.2). Since transport measurements showed no gaps in this doping range [26–28], we assign the broad background ΔR/R to the disruption of highly correlated liquid states from pump excitation across pseudo gaps. Thus, the observed ν = -4/3, -3/2, and -5/3 states are likely incipient, emerging from the highly correlated liquid states.

While the above experiments are carried out at a sample temperature of 2.0 K, we also perform temperature-dependent measurements (Extended Data Fig. 3-5). On the electron side, many of the charge order states exist at T = 7 K, with ν = 1 persisting to T = 20 K. On the hole side, the ν = -4/3, -3/2, and -5/3 states are no longer resolved at elevated temperatures, but there is evidence for pump-induced reduction in oscillator strength persisting to T ≥ 7 K. The ν = -1 state can be resolved at early pump-probe delays at T = 20 K, which is higher than the critical temperature of $T_c$ ~13 K reported earlier[1,2]. These findings suggest that our transient measurement is sensitive to not only correlated insulators but also to residual electronic correlations above nominal $T_c$.

**Time domain detection of fractional fillings of the Chern bands: θ = 3.1º**

We carry out transient measurements on a second tMoTe₂ sample (D2) with θ = 3.1º. While the competition between kinetic energy and Coulomb energies varies with θ, band topologies of tMoTe₂ are unchanged for the two twist angles (θ = 3.1º and 3.7º) [33–35]. Fig. 2a and Fig. 2b show ΔR/R maps at representative delays, Δt = 13 ps and 450 ps, respectively. The lower moiré unit cell density of D2 compared to D1 allows us to reach filling factors close to ν = ±3. While the abundant charge ordered states on the electron side remain clearly observed, those from the hole side are harder to resolve. To improve resolution, we present 2D spectral maps in derivative form, d(ΔR/R)/dn, where n is the doping density ($\propto V_g$). Figs. 2c show d(ΔR/R)/dn spectral maps averaged over Δt = 13, 25, and 45 ps; each state is identified at the inflection point connecting a peak and a valley. Fig. 2d shows selected line-cuts at Δt =13 ps and 450 ps. States at ν ≥ -1 are resolved in both time windows, while those between ν = -1 and -3 are only observed at Δt = 13 ps, not at Δt = 450 ps. The three states at ν = -4/3, -3/2, and -5/3 observed in D1 are reproduced in D2.

Between $\nu = -2$ and $-3$, we resolve three new states at $\nu = -8/3, -5/2$, and $-7/3$. As in D1, the $\nu = -4/3, -3/2, -5/3, -8/3, -5/2$, and $-7/3$ states in D2 all exhibit small features on top of intense and broad $\Delta R/R$ signal. These states are likely incipient from highly correlated liquids with pseudo gaps. In addition to the new states at $\nu < -1$, we observe a series of fractional states at $\nu = -1/5, -1/3, -1/2, -2/3, -4/5$. Overall, we resolve a total of 29 states.

**Control Sample at $\theta = 5.5°$**

As a control to the above findings of correlated states at $\theta = 3.7°$ and $3.1°$, we carry out measurements on a tMoTe$_2$ sample with a larger twist angle of $\theta = 5.5°$ (D3), Extended data Fig. 6. No correlated states have been found at such a large twist angle [1–4] and we confirm this in doping-dependent photoluminescence (PL) measurement. Consistent with static PL sensing, no states are detected in pump-probe spectral mapping in the entire doping range ($-5.5 \times 10^{12}$ to $5.5 \times 10^{12}$ cm$^{-2}$).

**Melting dynamics: electronic vs. phononic**

Having identified a zoo of states, we now investigate their melting and reordering dynamics. We present $\Delta R/R$ maps at $\Delta t = 1.5 – 300$ ps, Fig. 3 for D1. Similar spectral maps for D2 are shown in Extended Data Fig. 7. The correlated states are observed in the entire time window, with varying degrees of resolution. On the electron side, there are two distinct time windows of melting and recovery, as evidenced by the $\Delta R/R$ signal reaching the first minimum between 1.5 and 13 ps. Following partial recovery, $\Delta R/R$ decreases again on the longer time scale of ~300 ps, before recovering on ns time scales (see Fig. 4 for time-traces). On the hole side, we observe primarily the first melting characterized by a minimum at $\Delta t$ ~13 ps.

To understand the origins of melting/recovery processes, we present in Fig. 4 selected temporal profiles for D1 at (a) $\nu = \pm 1$, (b) $\nu = \pm 4/3$, (c) $\nu = \pm 3/2$, (d) $\nu = 0$ at the probe photon energy where $\Delta R/R$ reaches minimum. Similar profiles for other filling factors are summarized in Extended Data Fig. 8. For electron doping, the time profiles are characterized by initial decays with time constants of $\tau_{m1} = 2\text{-}3.5$ ps and recoveries with time constants of $\tau_{r1} = 20 – 40$ ps. This stage following excitation can be attributed to disordering from charge hopping [16–18] and we call this "electronic melting". As control, we confirm that this fast melting/recovery process is absent for $\nu = 0$, Fig. 4d, where only the decaying tail of an initial spike is observed; the latter likely originates from transient response of the graphite gates[16,17]. In sample D2, there are clearly resolved states near $V_g = 0$, making the precise assignment of charge neutrality difficult (see Methods). The state closest to $V_g = 0$ (arrow in Fig. 2b or 2e) is likely a hole-doped state at $-1/7 \leq \nu < 0$. Confirming this, we show that the electronic melting process is observed at 2 K but disappears at 70 K ($> T_c$), Extended Data Fig. 9. At this low doping level, electronic melting and recovery occur with time constants of $\tau_{m1} = 1.5 \pm 0.5$ ps and $\tau_{r1} = 3.4 \pm 0.9$ ps, respectively, both shorter than corresponding time constants at higher hole doping levels.

For $\nu = 0, 1, 4/3$, and $3/2$, we observe at $\Delta\tau > 7$ps high frequency oscillations that are known coherent phonon wavepackets launched at the graphite electrodes by the excitation pulse and arrived at tMoTe$_2$ with a time delay (Extended Data Fig. 10)[36,37]. For $\nu = 0$, we find that the excitation of phonons in tMoTe$_2$ decreases the oscillator strengths of both excitons and trions (Extended Data, Fig. 11), as shown by $\Delta R/R$ reaching a minimum at $\Delta t = 300 \pm 20$ ps, Fig. 4d. However, this change alone does not explain the observations at $\nu = 1, 4/3$, and $3/2$ (Fig. 4a-c), as their magnitudes in the decrease of $\Delta R/R$ are much larger than that at $\nu = 0$. Moreover, the occurrences of the second minima for $\nu = 1, 4/3$, and $3/2$ are time delayed from that of $\nu = 0$ by ~150 ps, likely reflecting the time for re-heating the electrons by phonons. Thus, the second minima in $\Delta R/R$ observed for $\nu = 1, 4/3$, and $3/2$ reveal an additional mechanism for pump-induced increases in effective dielectric constants specific of the correlated states, i.e, re-melting of these states. We refer to this stage as "phonon melting", similar to strain wave induced melting of CDW and semiconductor-to-metal transitions [38,39]. The phonon melting is responsible for the much-enhanced contrast in transient spectra for some of the correlated states at long time delays, Fig. 1 and Fig. 2.

We now turn to the melting and recovery dynamics of states at $\nu = -1, -4/3$, and $-3/2$. While the $\nu = -1$ state is well-resolved, the $\nu = -4/3$ and $-3/2$ states are identified only as weak features on a broad continuum. These states are likely incipient, emerging from a highly correlated liquid with pseudo gaps. The time profiles at $\nu = -4/3$ and $-3/2$ probe predominantly the dynamics of the underlying correlated liquid. The most obvious difference from electron doping is that phonon oscillations and the subsequent phonon melting/recovery are much reduced for hole doping at $\nu = -1$, and becomes negligible for $\nu = -4/3$ and $-3/2$. The negligible phonon perturbation for $\nu = -1$ is associated with an absence of spectral changes in exciton and trion sensing, in contrast to a time-dependent blue shift observed for the phonon-perturbed ν = 1 state, Extended Data, Fig. 12. Moreover, the electronic melting processes of states from hole doping are slower than those of corresponding electron doping (Fig. 4a-c). The melting and recovery time constants for ν = -1, -4/3, and -3/2 are $\tau_{m1}$ = 4.3-6.5 ps and $\tau_{r1}$ = 24-38 ps, repectively.

Here we present a tentative interpretation, Fig. 4f. The key feature of twisted TMD bilayers is the layer skyrmion texture of interlayer tunneling [33]. There are three distinct stacking regions at small twist angles – AA, XM, and MX. Interlayer tunneling is large in the AA regions but vanishes in the MX/XM regions. When a charge is located in the MX/XM stacking region, the wavefunction is made of non-bonding states at MX in one layer and XM in the other. For electron doping, the moiré conduction band derive primarily from the $Q$ point due to interlayer hybridization [32,40] and are delocalized spatially in zigzag stripes across the entire moiré unit cell [40]. This results in a large energy oscillation ($\delta_{ph}$) by the breathing phonon, which modulates the interlayer distance and, thus, hybridization. The hybridized and spatially delocalized nature also facilitates inter-site hopping ($t_h$) of holons and doublons, thus, accelerating melting and recovery dynamics. In contrast, for hole doped states (ν < 0), the uppermost moiré valence band near the $K$ points is localized to the MX or XM regions that form the honeycomb lattice [33–35]. The layer localized and nonbonding nature suggests that hole-doped states are insensitive to interlayer distance, leading to small $\delta_{ph}$. Holon/doublon hopping is inhibited between neighboring MX/XM sites and is also small between next-nearest neighbor sites, thus slowing melting and recovery dynamics. In the presence of sufficiently large displacement field, the hole doped states are polarized to a single layer [1–4]; Holon/doublon hopping $t_h$ may be increased and the modulation of displacement field by the breathing mode may also increase the magnitude of $\delta_{ph}$, as are observed in Fig 4e.

While the qualitative proposal based on spatial wavefunction variations provides a tentative explanation of the differing dynamics of electron and hole doped states, theories are needed to quantitative understand these differences. Moreover, the honeycomb MX/XM lattice is also related to the formation of topological states; whether the robustness of some of the hole doped states as compared to the electron doped ones is related to topology remains an open question.

**Discussions**

Our results establish the presence of the ν = -4/3, -3/2, -5/3, -7/3, -5/2, and -8/3 states from fractional fillings of the Chern bands in tMoTe$_2$. The discovery of these states that have evaded detection in prior experiments [1–4,26,27,41] underscores the superior sensitivity of our time-domain approach. These states are located in the phase space with small or no ferromagnetic order [27,28]. While their origins cannot be confirmed here, some of these fractional states are candidates for the exotic topological states predicted in recent theories [7–15]. Kwan et al. [7] proposed the ν = -4/3 state potentially as an abelian FTI consisting of two copies of ν = -2/3 in the K and K' valleys, stabilized by phonon coupling and/or sufficient nonlocal Coulomb screening. Chen et al.[42] suggested ν = -3/2 as a non-abelian FCI, but the lack of spin/valley polarization[27] argues against this proposal. May-Mann et al. [11] proposed half-integer FQSH edge states to form pairs of charged counter-propagating bosonic modes; our observed ν = -3/2 and -5/2 states might be related to these proposed edges states. We are not aware of theories on the ν = -5/3 and -7/3 states; an intriguing possibility is that they are FCIs, but fluctuations of weak magnetic order may make experimental verification difficult. The ν = -8/3 state is weakly magnetic[27] and potentially a topological state in the second moiré band. Further theoretical studies are warranted.

An intriguing question is why states at ν = -4/3, -3/2, -5/3, -7/3, -5/2, and -8/3 have not been resolved in prior transport measurements, that showed no evidence of gap formation at these fillings[26–28]. These states are likely incipient, emerging from highly correlated liquid states in the broad doping range (-3 < ν < -1.2). While the pump-probe approach can be sensitive to remanent correlation and pseudo gaps, transport measurements are not. Improved transport measurements are needed to isolate the correlated insulator responses from interfering effects of liquid-like states. One may also fully develop the correlated gaps from incipient states by sample engineering, e.g., reduction in disorder and control of dielectric screening, to allow the detection of quantized transport, particularly with edge resolution.

## METHODS

### Device Fabrication and Doping

Flakes of the van der Waals materials used to fabricate the heterostructure devices – graphite, h-BN, and monolayer MoTe$_2$ – were mechanically exfoliated onto oxygen-plasma cleaned Si/SiO$_2$ substrates and identified by optical contrast under an optical microscope. Atomic force microscopy (AFM) was used to determine h-BN thickness and confirm the flakes to be free of residue. As MoTe$_2$ was air sensitive, its exfoliation, as well as the rest of the device fabrication process, was performed in an argon filled glovebox with H$_2$O and O$_2$ concentrations less than 0.1 ppm. Before starting the transfer, the MoTe$_2$ monolayer was cut in half by an AFM tip. Standard dry transfer techniques were used to fabricate the heterostructure. First, the top gate was formed by picking up h-BN, the top graphite flake, the top gate h-BN dielectric, and a graphite grounding pin. Next, the first half of the MoTe$_2$ flake was picked up, the other half was rotated by the desired angle, and then picked up and placed down to form the moiré superlattice. Finally, the stack was placed down on a prepared backgate. This backgate consisted of an h-BN dielectric layer on top of a graphite flake, as well as gold contacts and wire bonding pads deposited using standard electron beam lithography and E-beam evaporation, allowing for electrical contact to both gates and to the grounding pin. The back gate was AFM-cleaned in contact mode before use. The finished heterostructure was placed onto the SiO$_2$/Si substrate (SiO$_2$ thickness: 285 nm for D1 and 90 nm for D2) by melting down the PC at ~170 °C, and the stamp polymer was dissolved in anhydrous chloroform in a glovebox environment for 5 minutes, completing the device. For D1, the top and bottom h-BN gate dielectrics are both 35nm. For D2, the top gate h-BN is 32nm and the bottom gate h-BN is 37nm. We estimate the thickness of graphite electrodes from optical contrast on the Si/SiO2 substrate. For D1, the top gate graphite is approximately 2-3 graphene layers, and the bottom is ~10 nm. For D2, the top is approximately two graphene layers and bottom ~2-3 nm.

**Determination of carrier density and twist angle**

The thickness of the bottom hBN layer in the tMoTe2 devices was measured by an atomic force microscopy (AFM). We calculated the geometrical capacitance per unit area between the gate and the sample using the formula $C_g = \frac{\varepsilon_{hBN}\varepsilon_0}{d_{hBN}}$, where $\varepsilon_{hBN}$=3.0 is the dielectric constant of h-BN [1,2], $\varepsilon_0$ is the vacuum permittivity, $d_{hBN}$ is the thickness of the top or bottom h-BN layer. This equation applies to either side of tMoTe2 sample to give top ($C_{tg}$) or bottom ($C_{bg}$) capacitance. The carrier density in tMoTe$_2$ was determined by $n = (V_{tg}C_{tg}+V_{bg}C_{bg})/e$, where $V_{tg/bg}$ is the top/bottom gate voltage, and $e$ is the elementary charge. The twist angles of the tMoTe$_2$ were determined from the spectral features observed in reflectance contrast measurements. The exciton resonance in tMoTe$_2$ shows abrupt changes at superlattice filling factors ν = ±1 (Fig. 1 and Fig. 2), enabling us to extract the corresponding carrier density. The twist angle was calculated from $\theta = \sqrt{\sqrt{3}/2} \cdot a_0 \cdot 180/\pi$, where $a_0$ = 3.52 Å is the MoTe$_2$ lattice constant.

**Reflection contrast measurements**

Each tMoTe$_2$ sample was mounted in a closed-cycle cryostat (Quantum Design, OptiCool) for all experiments, with a base temperature of 1.97 K during experiments. Steady-state reflectance measurements were carried out using a broadband lamp (Thorlabs, SLS201L). To minimize heating, a 1100 nm ±50 nm band-pass filter and a neutral density filter were employed. After collimation, the lamp light was focused onto the back pupil of the objective to illuminate the sample area using a 100X, 0.75 NA objective. The excitation power of the white light was maintained at approximately 100 nW. The reflected light was collected by the same objective, and a dual-axis galvo mirror scanning system was employed. Following the spatial scanning of the sample, controlled by the angles of the galvo mirrors, the reflected light was spatially filtered through a pinhole and then dispersed by a spectrometer onto an InGaAs array (PyLoN-IR, Princeton Instruments).

**Pump-probe measurements**

We carry out pump-probe experiments using femtosecond pulses (400 kHz, 1050 nm, 250 fs) generated by a solid-state laser (Light Conversion, Carbide). The laser output is split into two beams to form the pump and probe arms. For the probe, a portion of the fundamental beam is focused into a YAG crystal to generate a stable white light continuum, which is then spectrally filtered (1125 nm, 50 nm band-pass filter) to cover the exciton and trion energies in MoTe$_2$. The pump beam is directed to a motorized delay stage to control the pump-probe delay, Δt, and then focused onto a second YAG crystal to generate another broadband white light continuum, which is subsequently filtered down to 1250 ± 50 nm. After filtering, the pump beam passes through an optical chopper to generate alternating pump-on and pump-off signals. The pump and probe beams are then directed collinearly onto the sample through a 100X, 0.75 NA objective. The pump and probe spot diameters are approximately 1.5 μm and 1 μm, respectively. The pulse duration at the sample was estimated to be ~ 200 fs from the coherent artifact in pump-probe cross correlation. The same objective is used to collect the reflected light, which is spectrally filtered to remove the pump component and then dispersed onto an InGaAs detector array (PyLoN-IR, Princeton

Instruments). The pump-on and pump-off spectra at varying Δt are used to calculate the transient reflectance signal (ΔR/R), where ΔR = R(ΔT) – R and R is reflectance without pump. Note that, at the pump photon energy used here, there is no detectable excitation directly of the semiconducting MoTe$_2$. Such excitation would lead to Pauli blocking which would give rise to a nearly instantaneous bleaching signal at exciton/trion energies; this was not observe in time-resolved spectra. We use excitation photon energy below the optical gap of MoTe$_2$ to avoid the creation of excitons; pump excitation above the optical gap would result in overwhelming response of bleaching and recombination dynamics of excitons/trions. In the measurement, we vary the pump fluence from ρ = 7 to 42 μJ/cm$^2$, use a probe fluence of 22 μJ/cm$^2$ for all the measurement, Extended data fig. 13. The peak amplitude of ΔR/R signal is found to scale linearly with pump fluence. The electronic melting and recovery dynamics are independent of ρ in this range. In all data presented in the text, we use constant pump and probe fluences of 42 μJ/cm$^2$ and of 22 μJ/cm$^2$, respectively.

**Coherent phonons launched at graphite electrodes**

The few-layer graphene/graphite gate electrodes in a vdW structure are known to function as opto-elastic transducers [36,37]. Since the density of electrons that can be excited by the pump photon of $\hbar\omega$ = 0.99 eV is approximately three orders of magnitude higher in the graphite electrode than that in the doped moiré structure, most of the absorption of pump light occurs in the former. The resulting strain field in the photo-excited graphite electrodes launch acoustic phonon wavepackets that propagate through the h-BN spacer and reach the tMoTe$_2$ at a time delay determined by the spacer thickness and the phonon group velocity. We estimate that phonon group velocity from the arrival time of the wavepackets and the h-BN thickness: $v_{ph}^g \approx d_{hBN}/t_{delay} \approx$ 3.5 km/s, in agreement with the longitudinal acoustic phonon velocity in h-BN [37]. The phonon wavepackets modulate the electronic structure and the effective dielectric environment of tMoTe$_2$, enabling detection of the phonon oscillations through coherent changes in the exciton resonances in MoTe$_2$. The acoustic phonon wavepackets launched from the graphite electrodes are broad band and those detected at tMoTe$_2$ are frequency selected by resonant modes, particularly inter-layer breathing (LB) mode in the multilayer vdW structure [36,37].

**Assignment of Filling Factors**

The filling factor ν is defined as the number of electrons or holes per moiré superlattice site. To determine ν, we use the well-established insulating states, ν = 0, ν = ±1 and ν = ±2 as reference points to extract the conversion factor between gate voltage ($V_g = V_{tg} + V_{bg}$) and the filling factor for both electron and hole doping. We start by identifying the gate voltages for all insulating states, determined by the maximum change in reflection observed in the pump-probe map (e.g., Fig. 1g). For distinct and closely spaced states, we determine their peak positions and full width at half-maximum (FWHM) by fitting Lorentzian profiles to the exciton/trion spectral weight as a function of gate voltage. To reduce interference effects from the top and bottom hBN layers and the substrate, which can distort the reflection spectrum, we integrate the reflection contrast over an energy window around the exciton/trion peak to extract the spectral weight (see Extended Data Fig. 12). The fitting results for all states, along with the corresponding gate voltages, are presented in Extended Data Table 1.

For device D1, the gate voltages corresponding to ν = 0, ν = 1 and ν = -1 are determined from the static and transient reflection. Using a linear fit to these points, we establish the relationship between filling factor and gate voltage. The filling factors for other insulating states are then calculated based on their gate voltages and the conversion factor. Each state is assigned to the closest rational number with a small denominator, as these generally represent lower-energy states.

For device D2, we applied the same method to determine the states between ν = 1 and ν = -1. Two peaks close to ν = 0 (|ν| < 1/7), were not assigned specific filling factors due to the limited resolution and uncertainty in the precise gate voltage at ν = 0. For states |ν| >1, the gate voltage becomes more effective; therefore, we used ν = -2, ν = -1 and ν = 0 to define the relationship for -2 < ν < -1 states and use ν = 0, ν=1 and ν =2 to define the relationship for ν >1. For less well-developed states (ν < −2), we estimated the peak positions and widths from the first derivative of the reflection contrast contour plot with respect to energy (see Fig. 2c and d for the averaged derivative map at different time delays, and Extended Data Fig. 7 for individual time delays). The gate voltages corresponding to these correlated states are determined by the inflection points in the derivative map and the associated spectral weight.

**ACKNOWLEDGEMENTS**

Time-resolved spectroscopy experiments were supported by Programmable Quantum Materials, an Energy Frontier Research Center funded by the U.S. Department of Energy (DOE), Office of Science, Basic Energy Sciences (BES), under award DE-SC0019443. Methodology development for the pump-probe experiments was support by the DOE-BES under award DE-SC0024343 (XYZ). The device fabrication was supported by DOE-BES under the award DE-SC0018171(XX), using the facilities and instrumentation supported by NSF MRSEC DMR-2308979. XYZ and XR acknowledge partial support by the US Army Research Office, grant number W911NF-23-1-0056, for supporting the temperature-dependent experiments. XYZ acknowledges support by the Department of Defense (DOD) Multidisciplinary University Research Initiative (MURI) under grant number W911NF2410292 for the development of mechanistic models of electron-phonon coupling. YW acknowledges the Max-Planck New York Center (MPNYC) for fellowship support. EAA acknowledges support from the Simons Foundation as a Junior Fellow in the Simons Society of Fellows (965526). Facilities supported by the Materials Science and Engineering Research Center (MRSEC) through NSF grant DMR-2011738 were utilized in this work. K.W. and T.T. acknowledge support from the JSPS KAKENHI (Grant Numbers 21H05233 and 23H02052) and World Premier International Research Center Initiative (WPI), MEXT, Japan. We thank Nishchhal Verma and Daniel Munoz Segovia for helpful discussions, Yinjie Guo, Jordan Pack, and Sanat Ghosh for assistance with sample preparation.

**Author Contributions**

Y.W., X.Y.Z., and X.X. conceived this work. Y.W., along with J.C., conducted all spectroscopic measurements, analyzed, and interpreted the results, with the assistance of E.A.A. and YL, and inputs from D.B., X.R., and J.C.H. E.A. was responsible for the fabrication and characterization of sample (D1) and W.L. for sample (D2), under the supervision of X.X. T.T. and K.W. provided the hBN crystal. J.I., R.Q., X.H., and D.X. contributed to mechanistic interpretations. The manuscript was prepared by Y.W. and X.Y.Z., incorporating inputs from all coauthors. X.Y.Z. supervised the project. All authors read and commented on the manuscript.

**Competing Interests.** The authors declare no competing interests.

**Data Availability Statement.** The data within this paper are available upon reasonable request.

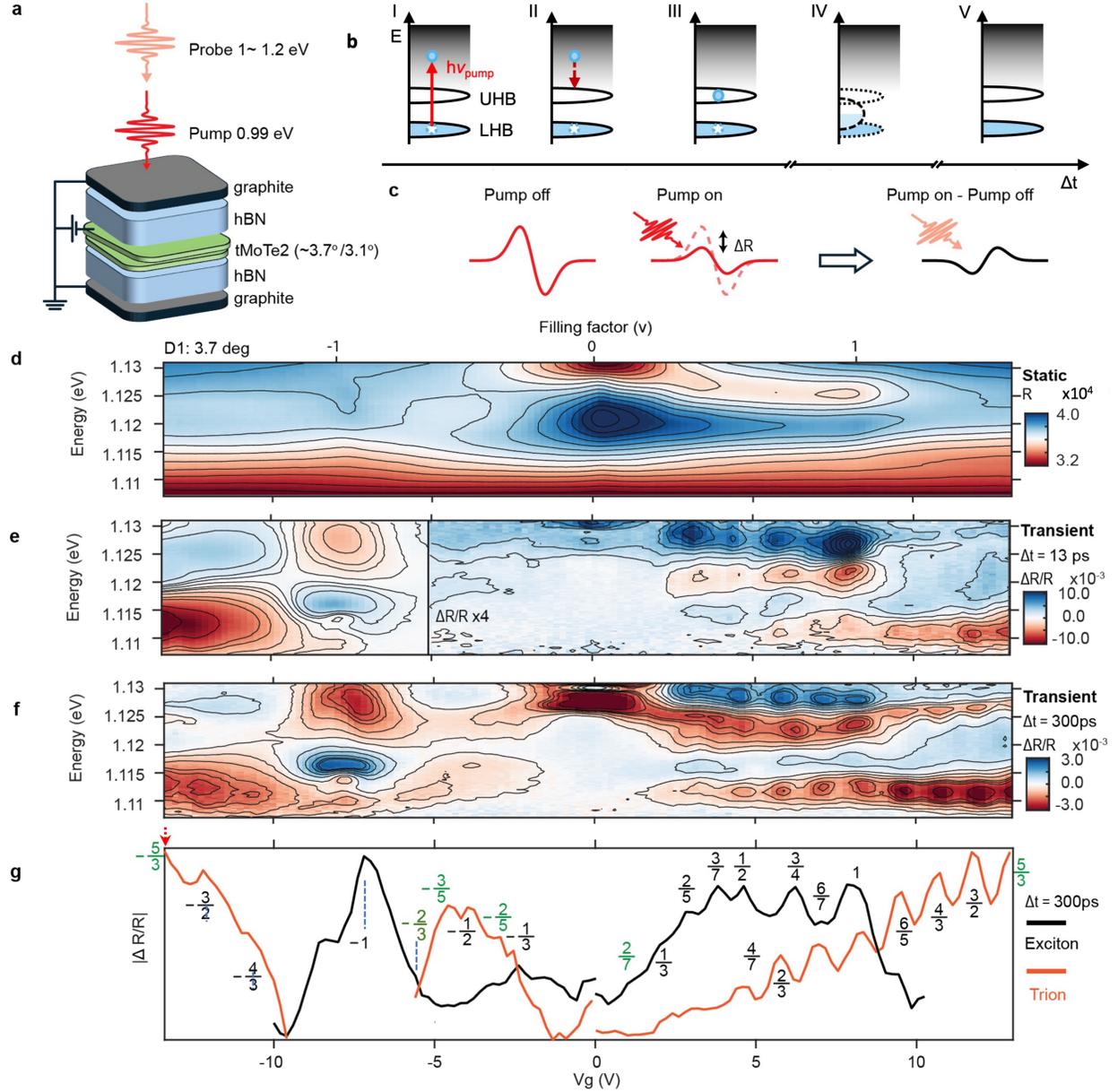

**Figure 1. Pump-probe spectroscopy detects hidden states at fractional fillings in tMOTe$_2$.** **a.** Schematics of a dual-gated MoTe$_2$ bilayer device (D1) with a twist angle $\theta = 3.7°$. The pump photon energy is $\hbar\omega_1 = 0.99$ eV and the broadband probe with photo energies $\hbar\omega_2 = 1.0$-$1.2$ eV. **b.** Schematic illustration of pump-induced processes in a correlated state, illustrated here for a Mott state with a lower and upper Hubbard band (LHB and UHB). Excitation across the correlated gap (I) creates a hot carrier, which relaxes on ultrafast time scales (II) to form a holon-doublen pair across the gap (III). These are followed by disordering/melting as represented by a reduction or closing of the correlated gap (IV), followed by recovery on a longer time scale (V). **c.** Schematic illustration of an exciton or trion reflection spectrum from a correlated state, with change induced by the pump and the corresponding differential reflectance spectrum. **d.** Static reflection spectrum ($R$) as a function of gate bias ($V_g$) and probe photon energy ($\hbar\omega_2$). **e. & f.** Transient reflection spectra, $\Delta R/R$ (pseudo color), where $\Delta R = R(\Delta t) - R$, as a function of $V_g$ and $\hbar\omega_2$ for delay times of $\Delta t = 13$ ps and 300 ps, respectively. **g.** Linecuts from the spectral maps **f** at exciton (black) and trion (red) energies. The calibrated filling factors (ν) are indicated on the line cuts. All experiments carried are out at a sample temperature of T = 2.0 K.

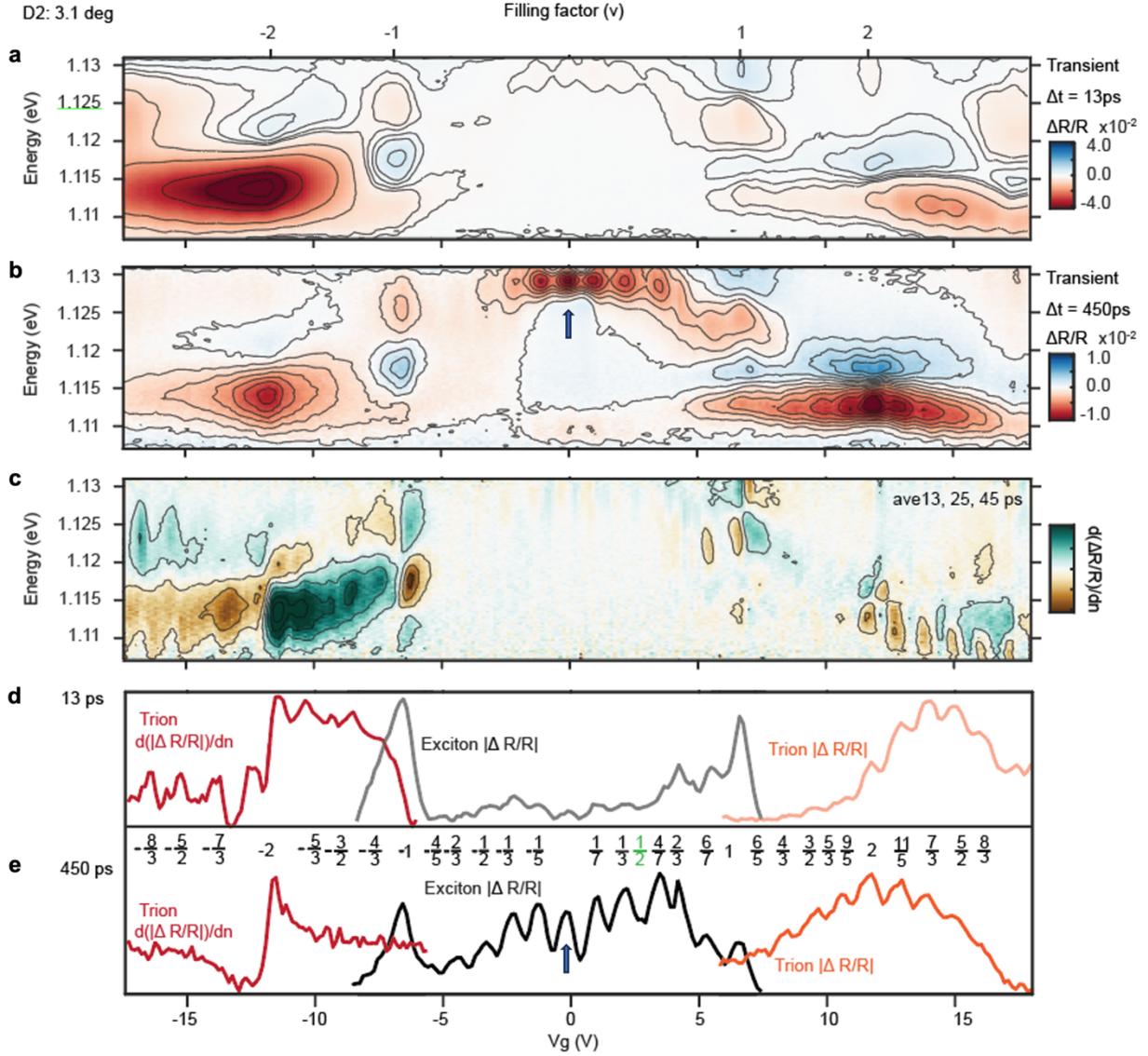

**Figure 2. Pump-probe spectroscopy detects hidden states at fractional fillings of the first and second Chern bands in tMOTe$_2$ device 2 (D2) with a twist angle θ = 3.1°.** Transient reflection spectra, $\Delta R/R$ (pseudo color), where $\Delta R = R(\Delta t) - R$, as a function of $V_g$ and $\hbar\omega_2$ for delay times of $\Delta t$ = 13 ps **(a)** and 450 ps **(b)**, respectively. Panels **(c)** is a derivative spectral map d($\Delta R/R$)/dV$_g$, averaged over three early delays times ($\Delta t$ = 13, 25, and 45 ps). The color scales are normalized, negative (blue) to positive (red). (d) Linecuts from the spectral maps of $\Delta R/R$ at exciton (black or grey), trion (red) energies, and from derivative spectral map d($\Delta R/R$)/dn. The calibrated filling factors (ν) are indicated on the line cuts.). All experiments carried are out at a sample temperature of T = 2.0 K.

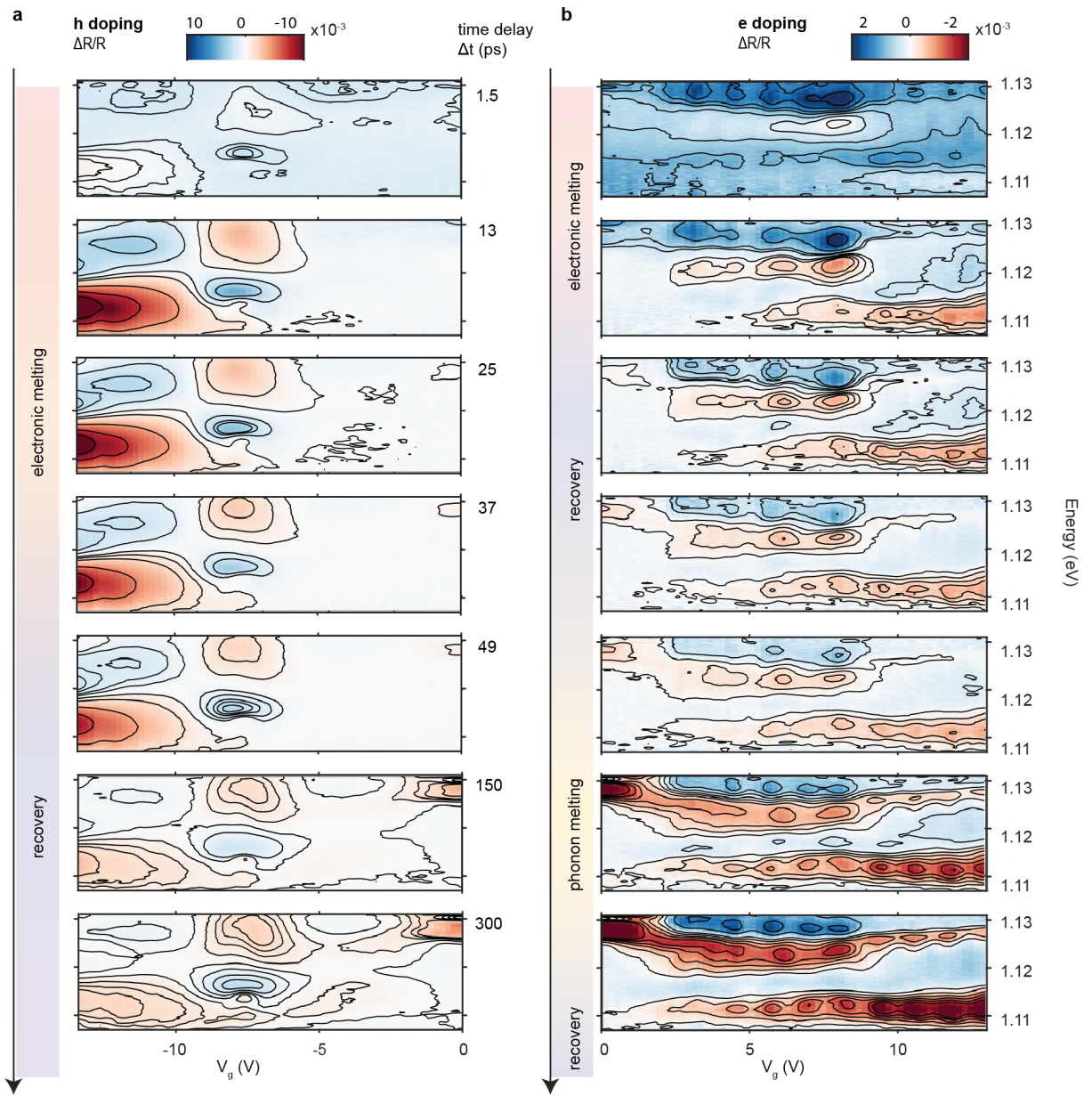

**Figure 3. Melting and recovery dynamics of correlated states**. Transient reflection spectra at the indicated pump-probe delay times (from top to bottom) of $\Delta t$ = 1.5, 13, 25, 37, 49, 150, 300 ps for **a.** hole doping and **b.** electron doping. The arrows illustrate the time windows for electronic melting/reordering and phonon melting/reordering. All spectra obtained from Device D1 ($\theta = 3.7°$), at a sample temperature of T = 2.0 K.

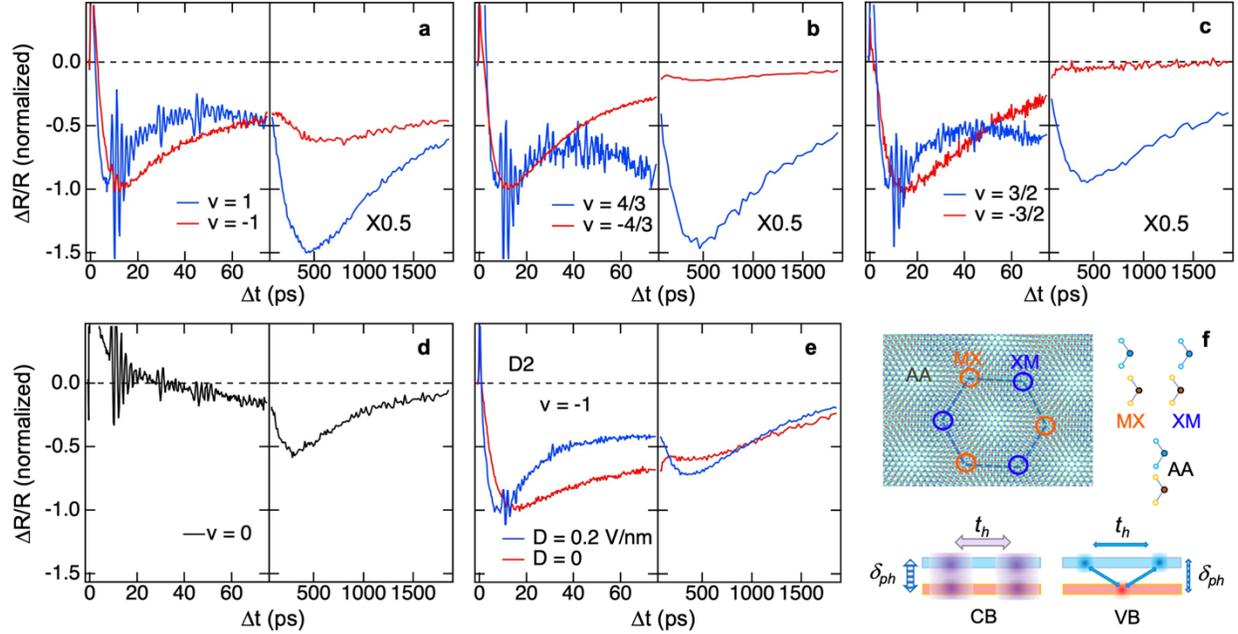

**Figure 4. Melting and recovery dynamics**. Time profiles of transient reflection (ΔR/R) for: **a** ν = ±1, **b** ν = ±4/3, **c** ν = ±3/2, and **d** ν = 0 from device D1 (θ = 3.7º). **e.** Time profiles of the ν = -1 state in device D2 (θ = 3.1º) at two displacement fields, D = 0.0 (red) and 0.2 V/nm (blue). Each profile is obtained at the probe photon energy where ΔR/R reaches minimum, integrated over a small spectral window (±2 meV). Note that each panel is divided into two scales (0-75 ps and 75 – 1875 ps). In panels **a**, **b**, **c**, and **e**, the ΔR/R values are normalized to the minima on the short time scale (< 20 ps). In panel **d**, there is no minimum on the short time scale (< 20 s) and the normalization factor was an average of those in panels **a-c**. The normalized ΔR/R values between 75-1875 ps in panels **a**, **b**, and **c** are scaled by a factor of 0.5. All data obtained at a sample temperature of T = 2.0 K with zero displacement field. **f.** Illustration of the moiré superlattice showing the layer-localized and non-bonding XM/MX (blue/orange) regions that form the honeycomb lattice, and the interlayer hybridized AA regions. The lower panel illustrates inter-site hopping ($t_h$) and modulation of electronic states by the breathing phonon ($\delta_{ph}$) in the conduction band (CB) and valence band (VB).

# Hidden States and Dynamics of Fractional Fillings in tMoTe$_2$ Moiré Superlattices

Yiping Wang[1,2], Jeongheon Choe[1], Eric Anderson[3], Weijie Li[3], Julian Ingham[4], Eric A. Arsenault[1], Yiliu Li[1], Xiaodong Hu[5], Takashi Taniguchi[6], Kenji Watanabe[7], Xavier Roy[1], Dmitri Basov[4], Di Xiao[5], Raquel Queiroz[4], James C. Hone[2], Xiaodong Xu[3,5], X.-Y. Zhu[1,1]

[1] Department of Chemistry, Columbia University, New York, NY 10027, USA

[2] Department of Mechanical Engineering, Columbia University, New York, NY 10027, USA

[3] Department of Physics, University of Washington, Seattle, WA 98195, USA

[4] Department of Physics, Columbia University, New York, NY 10027, USA

[5] Department of Materials Science and Engineering, University of Washington, Seattle, WA 98195, USA

[6] Research Center for Materials Nanoarchitectonics, National Institute for Materials Science, 1-1 Namiki, Tsukuba 305-0044, Japan

[7] Research Center for Electronic and Optical Materials, National Institute for Materials Science, 1-1 Namiki, Tsukuba 305-0044, Japan

---

[1] To whom correspondence should be addressed. E-mail: xyzhu@columbia.edu.

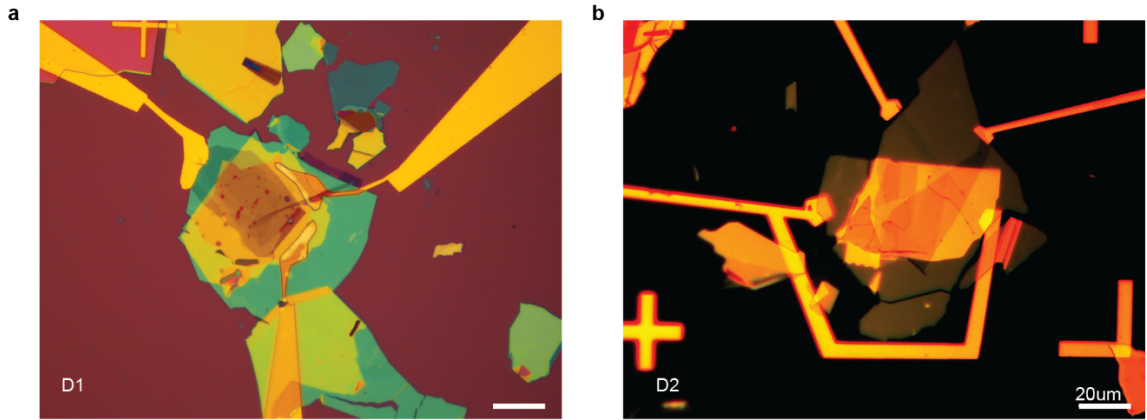

**Extended Data Fig. 1: Device Images. a-b.** 50X microscope image of device D1 and D2, the scale bar is 20 μm.

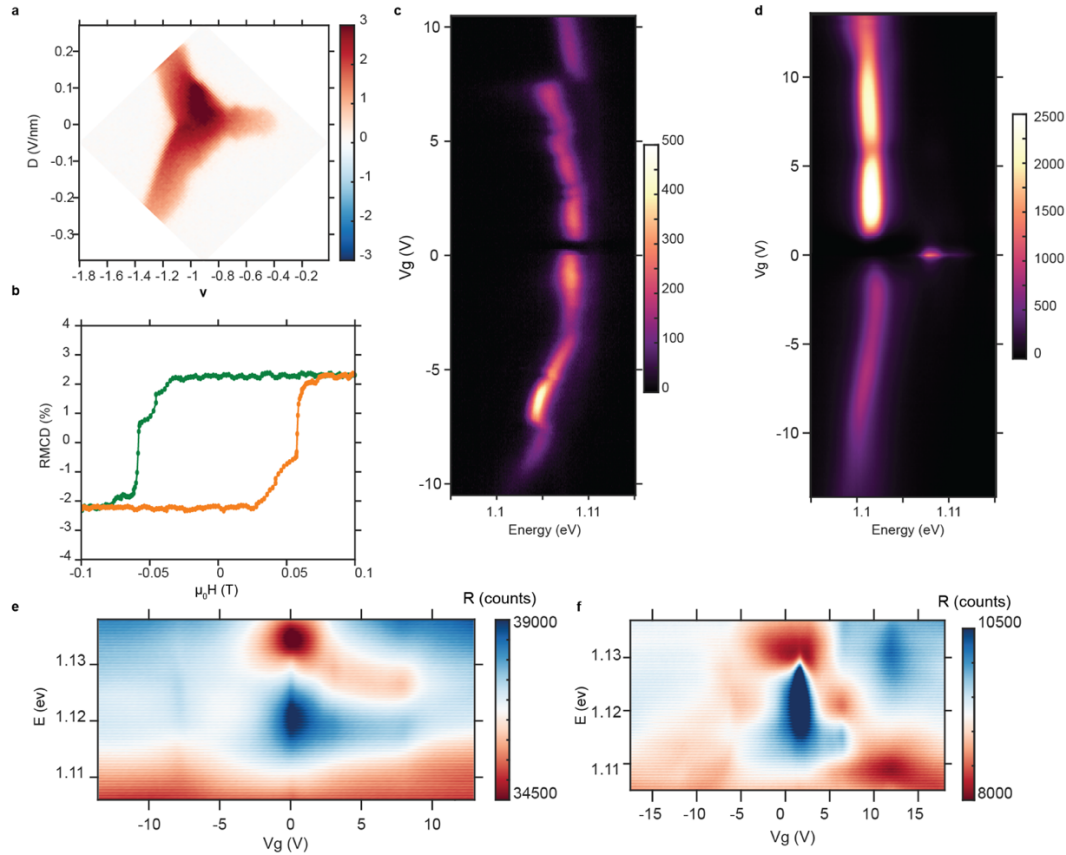

**Extended Data Fig. 2: Device characterization. a.** RMCD signal versus $v$ and perpendicular electric field $D$ at zero magnetic field $\mu_0H = 0$ (D1). The phase space with non-vanishing signal corresponds to the ferromagnetic state. **b.** RMCD signal versus vertical magnetic field with $\mu_0H$ swept back and forth at n = -0.4*10$^{13}$cm$^{-2}$ and D = 0V/nm (D2). **c.** D1: PL intensity plot as a function of doping and photon energy. **d.** D2: PL intensity plot as a function of doping and photon energy. **e.** Reflection as a function of doping and photon energy (D1). **f.** Reflection as a function of doping and photon energy (D2).

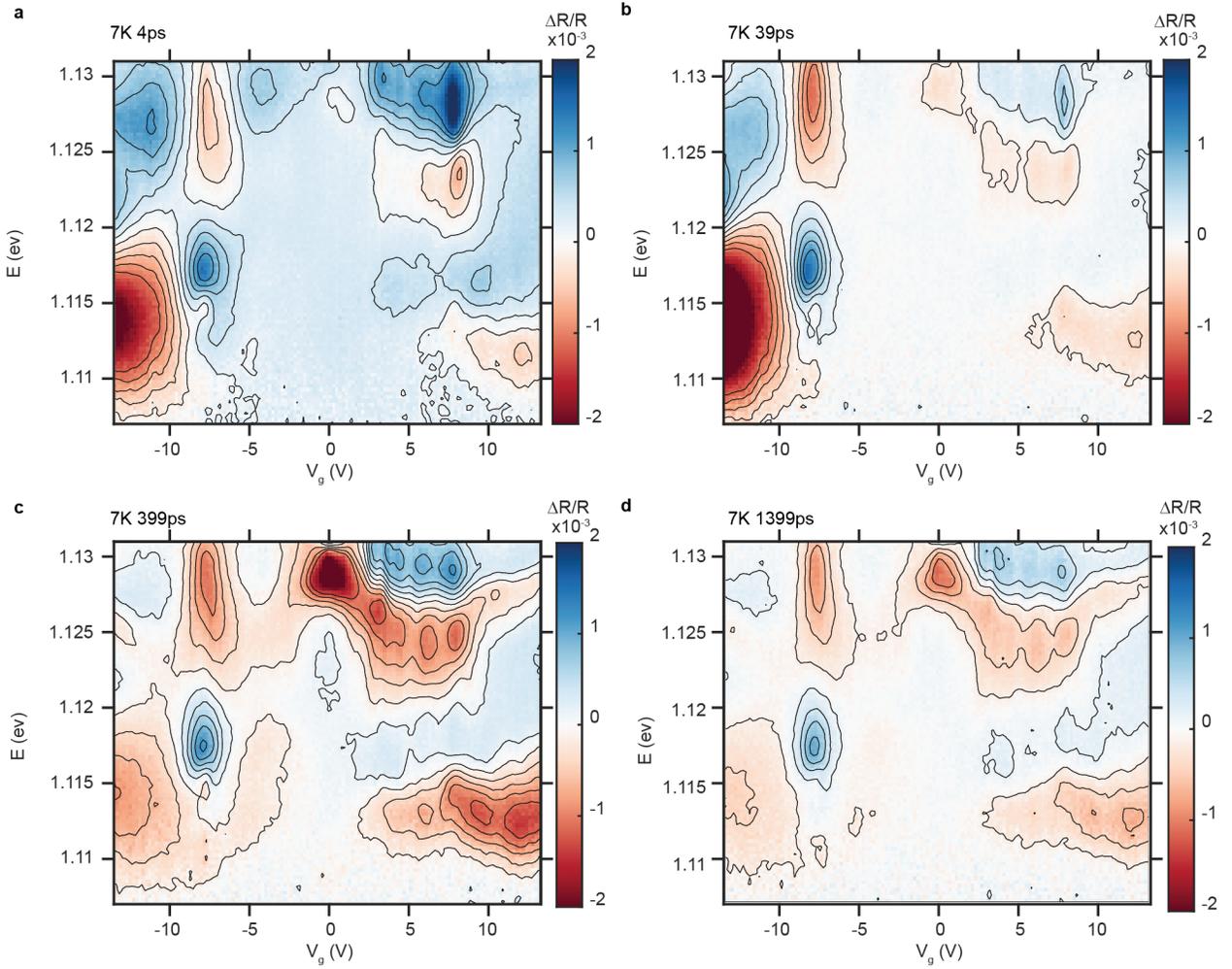

**Extended Data Fig. 3: Pump-probe spectroscopy at T = 7 K of tMoTe$_2$ device 1 (D1) with a twist angle θ = 3.7°.** Transient reflection spectrum as a function of carrier density (n) and probe photon energy E ($\hbar\omega_2$) at pump-probe delays of **a.** Δt = 4 ps, **b**. Δt = 39 ps, **c**. Δt = 399 ps and **d**. Δt = 1399 ps at 54 uJ/cm$^2$ (ΔR/R$_0$ color range -2x10$^{-4}$ to 2x10$^{-4}$). There is no external magnetic field (B = 0) or displacement field (D = 0).

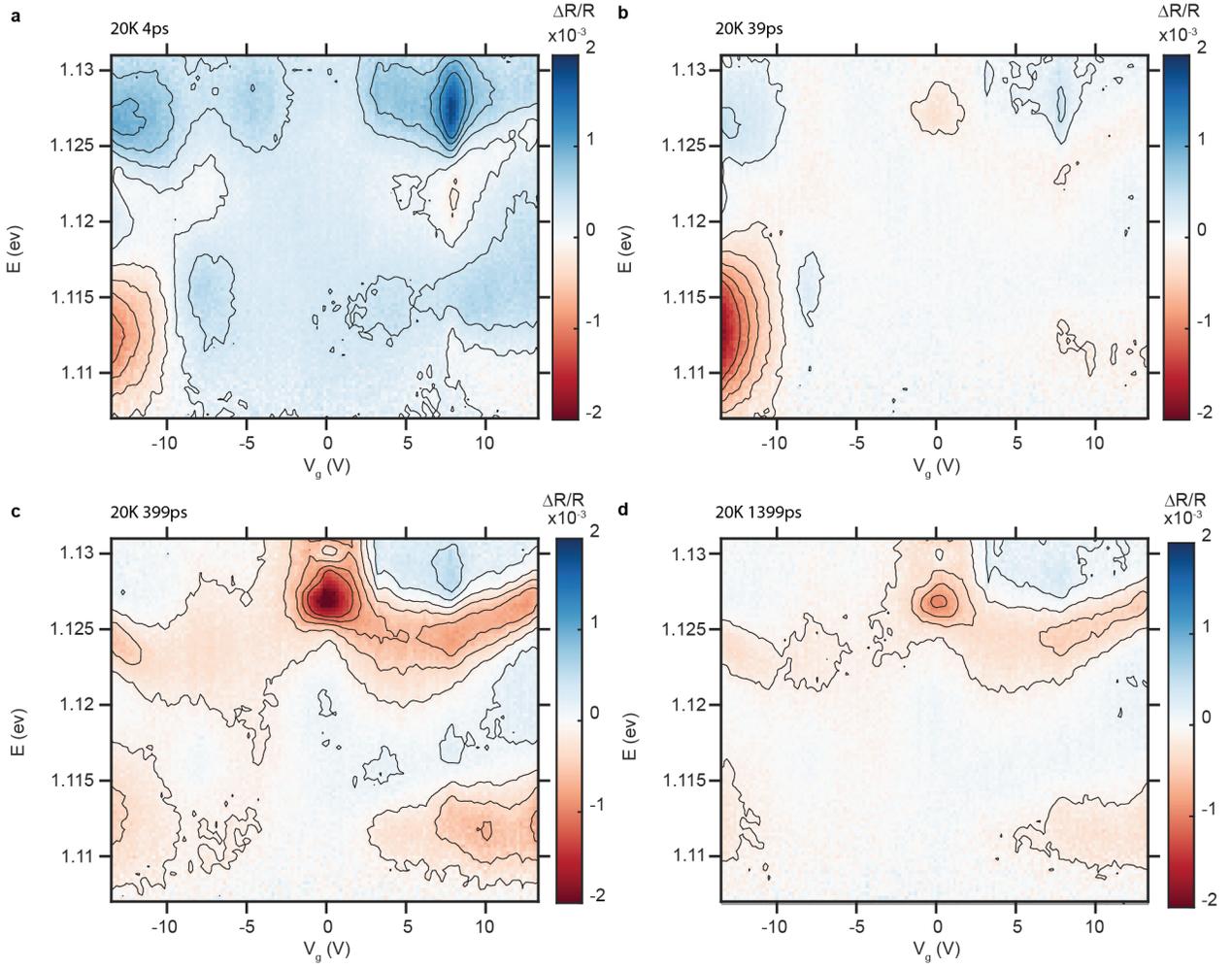

**Extended Data Fig. 4: Pump-probe spectroscopy at T = 20 K of tMoTe$_2$ device 1 (D1) with a twist angle θ = 3.7º.** Transient reflection spectrum as a function of carrier density (n) and probe photon energy E ($\hbar\omega_2$) at pump-probe delays of **a.** Δt = 4 ps, **b**. Δt = 39 ps, **c**. Δt = 399 ps and **d**. Δt = 1399 ps at 54 uJ/cm$^2$ (ΔR/R$_0$ color range -2x10$^{-4}$ to 2x10$^{-4}$). There is no external magnetic field (B = 0) or displacement field (D = 0).

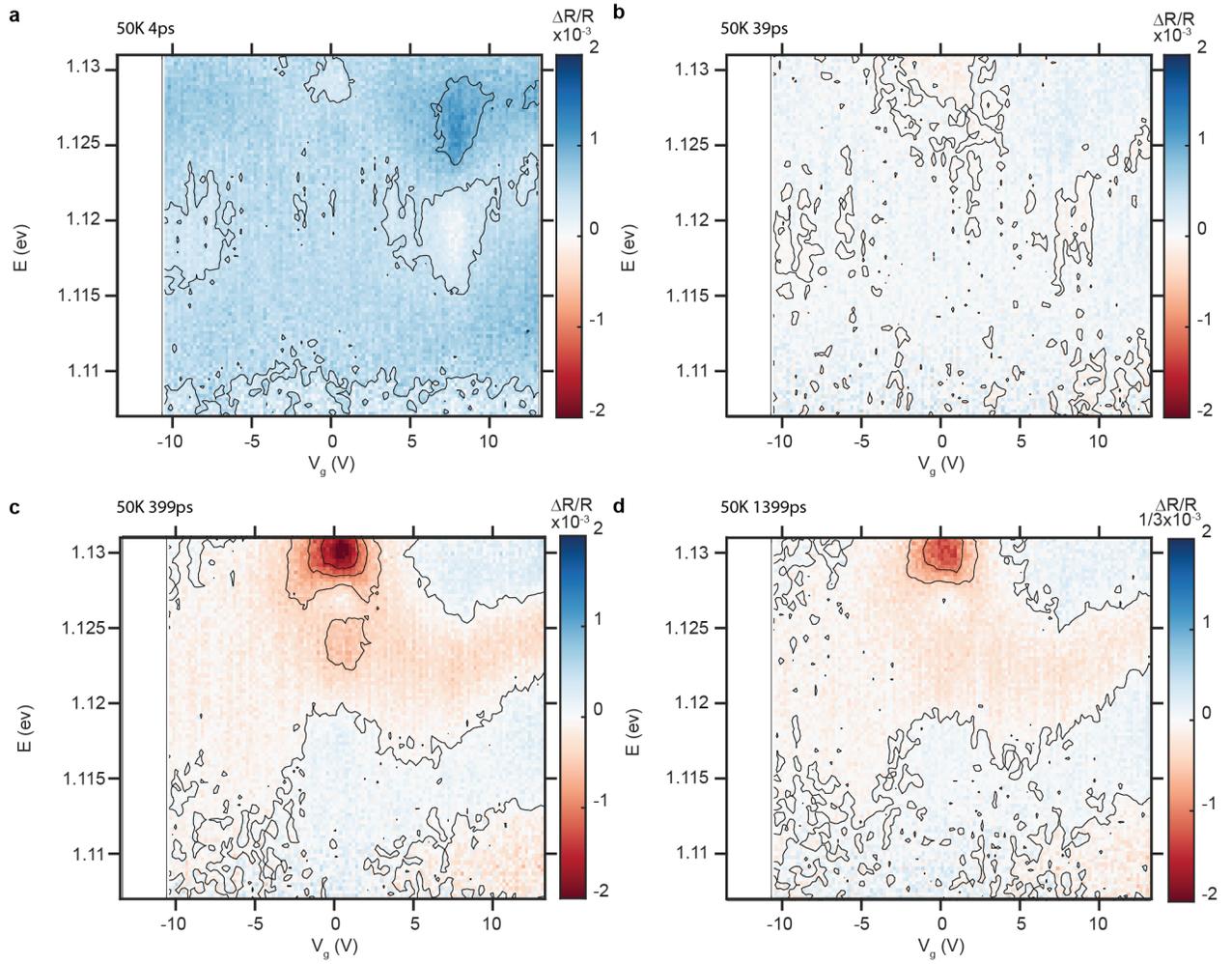

**Extended Data Fig. 5: Pump-probe spectroscopy at T = 50 K of tMoTe$_2$ device 1 (D1) with a twist angle θ = 3.7º.** Transient reflection spectrum as a function of carrier density (n) and probe photon energy E ($\hbar\omega_2$) at pump-probe delays of **a.** Δt = 4 ps, **b**. Δt = 39 ps, **c**. Δt = 399 ps and **d**. Δt = 1399 ps at 54 uJ/cm$^2$ (ΔR/R$_0$ color range -2x10$^{-3}$ to 2x10$^{-3}$). There is no external magnetic field (B = 0) or displacement field (D = 0).

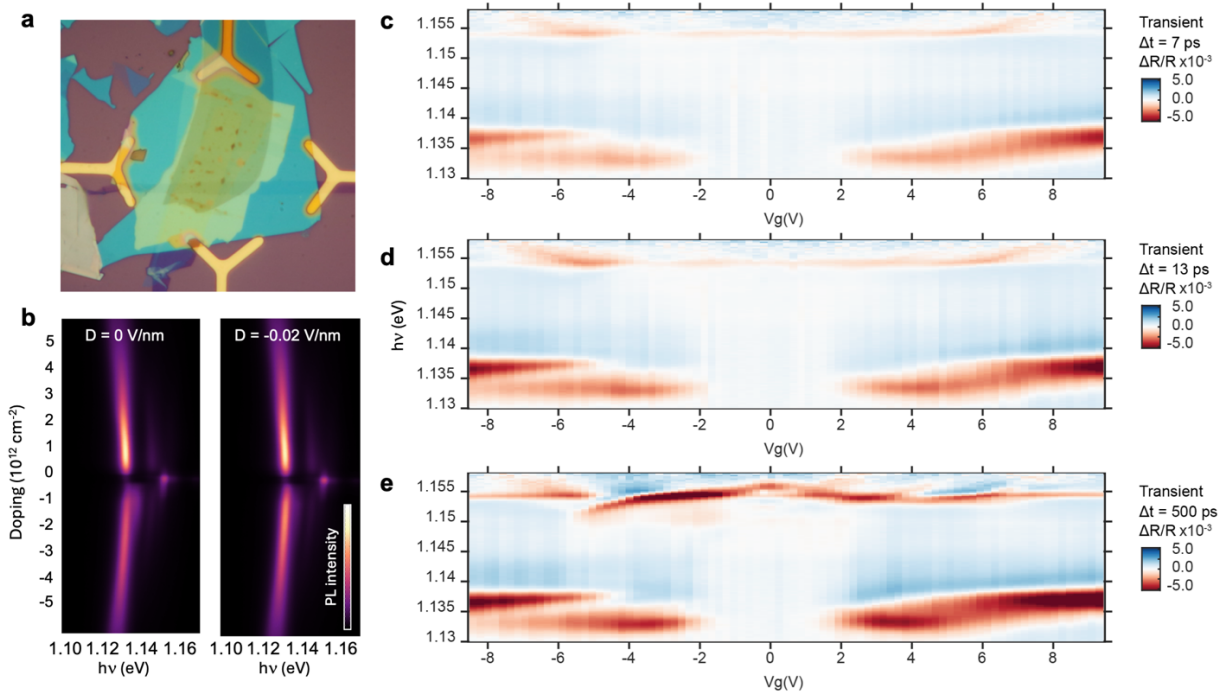

**Extended Data Fig. 6: Pump-probe spectroscopy of a tMoTe$_2$ device (D3) with a twist angle θ = 5.5°.** (a) Optical image (~100 μm x 100 μm) of device D3. (b) PL spectra as a function doping level (estimated from calculated capacitance). The left and right show displacements fields of D = 0 and -0.02 V/nm; the latter is used to compensate for a small build-in potential; (c-e) Transient reflection spectrum as a function of total gate voltage $V_g$ (V) and probe photon energy E ($\hbar\omega_2$) at pump-probe delays of **c.** Δt = 7 ps, **d.** Δt = 13 ps, and **e** Δt = 500 ps. The pump (0.99 eV) and probe fluences are at 42 μJ/cm$^2$ and 22 μJ/cm$^2$, respectively. Sample temperature T = 1.6 K. There is no external magnetic field (B = 0) or displacement field (D = 0). The gate voltage ranges in (c-e) correspond to the same estimated doping range shown in the PL maps (b).

No specific states (at particular $V_g$ or doping levels) are resolved in the pump-probe spectral maps at all three selected Δt values (7, 13, 500 ps), in agreement with the PL maps in (b). The transient spectral maps feature exciton and trion resonances, with energy splitting in certain doping ranges; these splitting features have been observed before and attributed to exciton/trion fine structures in MoTe$_2$ monolayers (*Nature Nanotech.* **2013**, *9*, 634-638; *Nature Nanotech.* **2017**, *12*, 144-149) and/or exciton polarons in tMoTe$_2$ bilayers (*Nature Nanotech.* **2022**, *17*, 934-939). The broad contrasts in exciton/trion ΔR/R signal vary slowly with Δt and likely result from the dynamics or hot-carrier relaxation, carrier-phonon scattering, phonon cooling, and balances in exciton and trion populations.

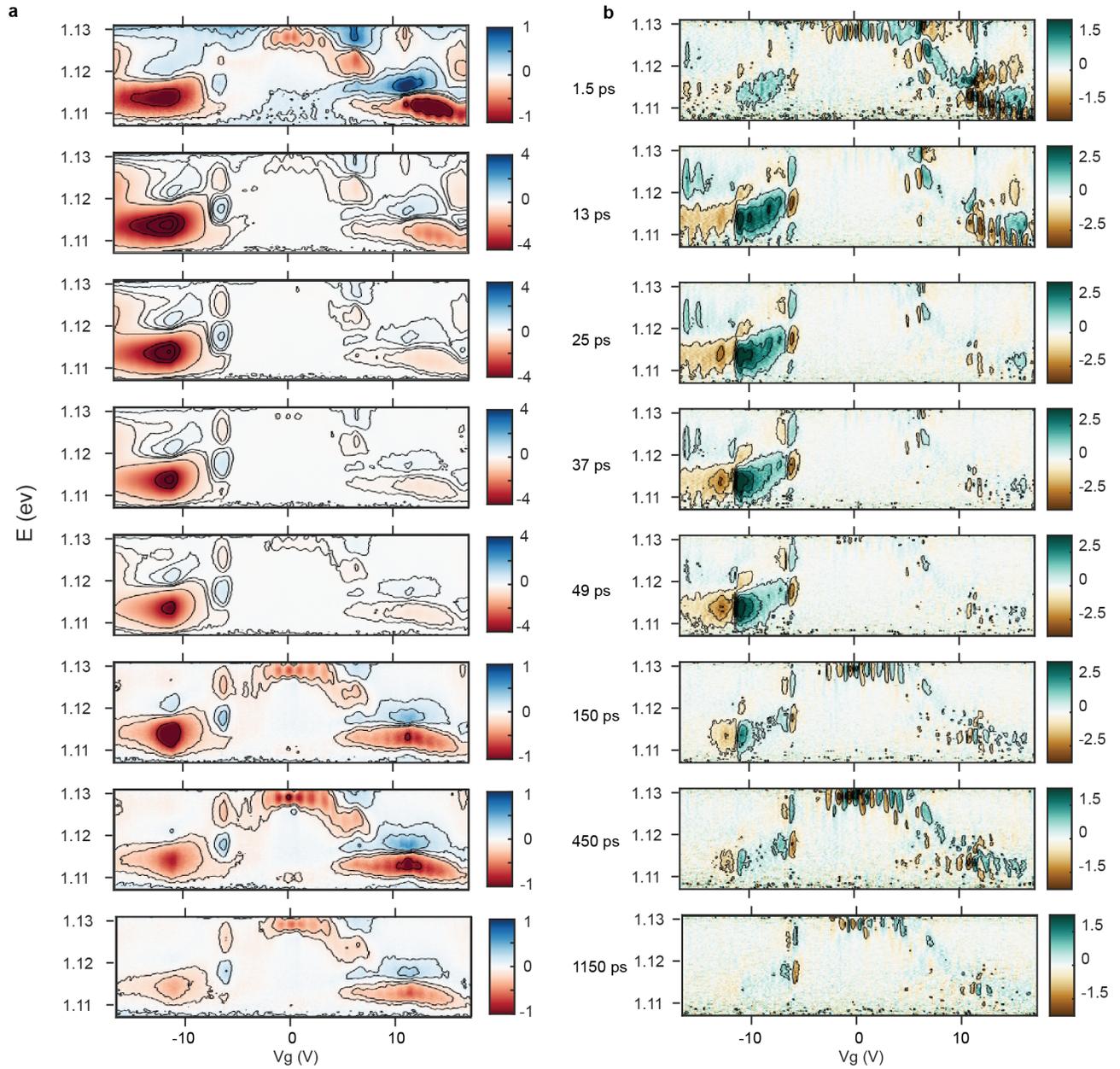

**Extended Data Fig. 7: Melting dynamics of correlated states**. **a.** Transient reflection spectra at the indicated pump-probe delay times (from top to bottom) of Δt = 1.5, 13, 25, 37, 49, 150, 450 and 1150 ps for **a.** hole doping and **b.** The corresponding 1$^{st}$ derivative (with respect to n) of the transient reflection spectra in **a**. All spectra obtained from tMoTe$_2$ device D2 (θ = 3.1°), at a temperature of T = 2 K, with no external electric or magnetic field.

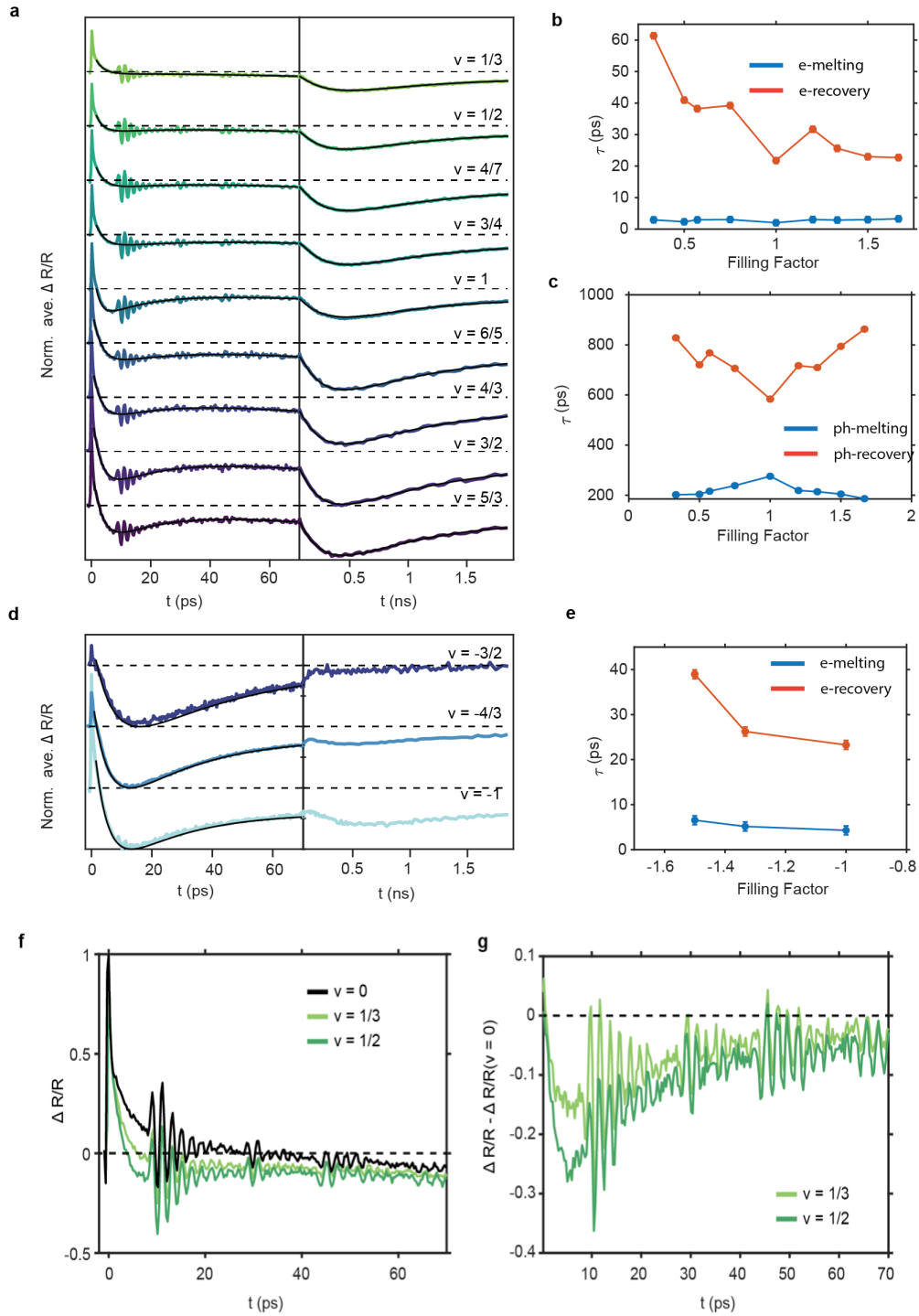

**Extended Data Fig. 8: Time profiles of difference states. a**. Time profiles of all the electron doping states in D1. **b.** electronic meting and recovery time constant of all the electron doped states. **c.** phonon melting and recovery time constant of electron doped states **d**. Time profiles of hole doped states. **e.** electronic meting and recovery time constant of all the hole doped states. **(f)** Time profile of v = 0, 1/3 and 1/2, data normalized to t = 0 ps. **(g)** Time profile difference of v = 1/3,1/2 to v = 0. All data obtained at a temperature of T = 2 K, with no external electric or magnetic field.

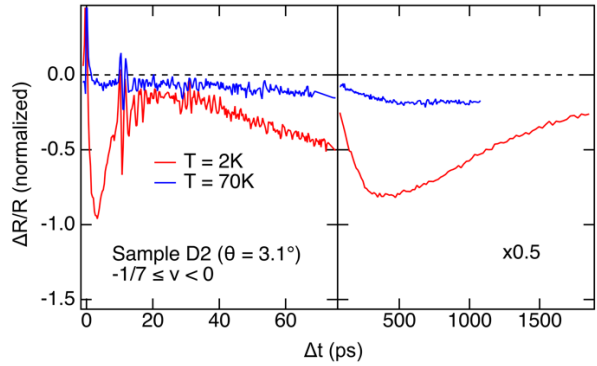

**Extended Data Fig. 9: Time profiles of a hole doped state of -1/7 ≤ ν < 0, which is closest to $V_g = 0$.** Upper: sample temperature T = 2 K (nominal reading on sample stage 1.58 K); Lower: sample temperature T = 70 K, which is above the $T_c$. Electronic and phonon melting processes are observed at 2 K, but not 70 K.

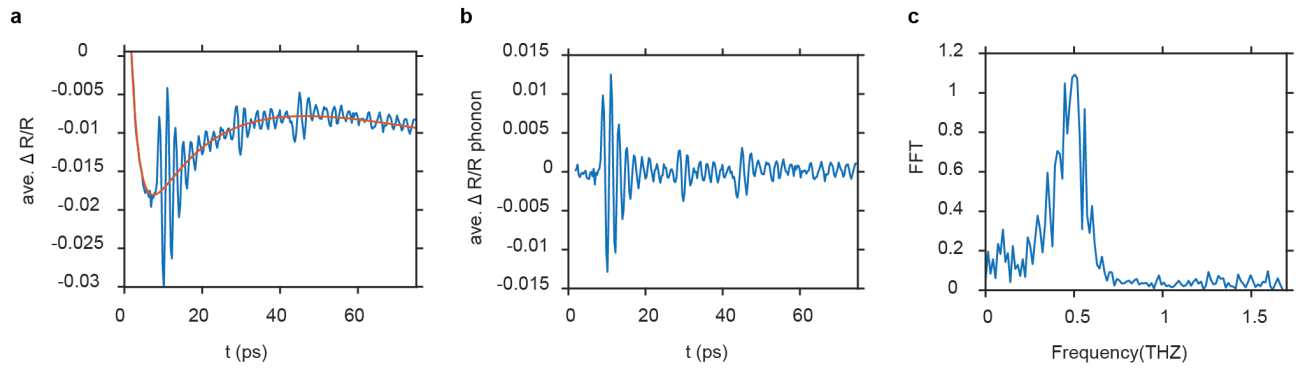

**Extended Data Fig. 10: Delayed arrival of coherent phonon wavepackets launched at the graphite electrodes. a.** Time profiles of the ν = 1 state with melting and recovery fitting (red). **b.** coherent phonon oscillation after the melting and recovery background is subtracted. **c.** Fourier transform of the coherent phonons.

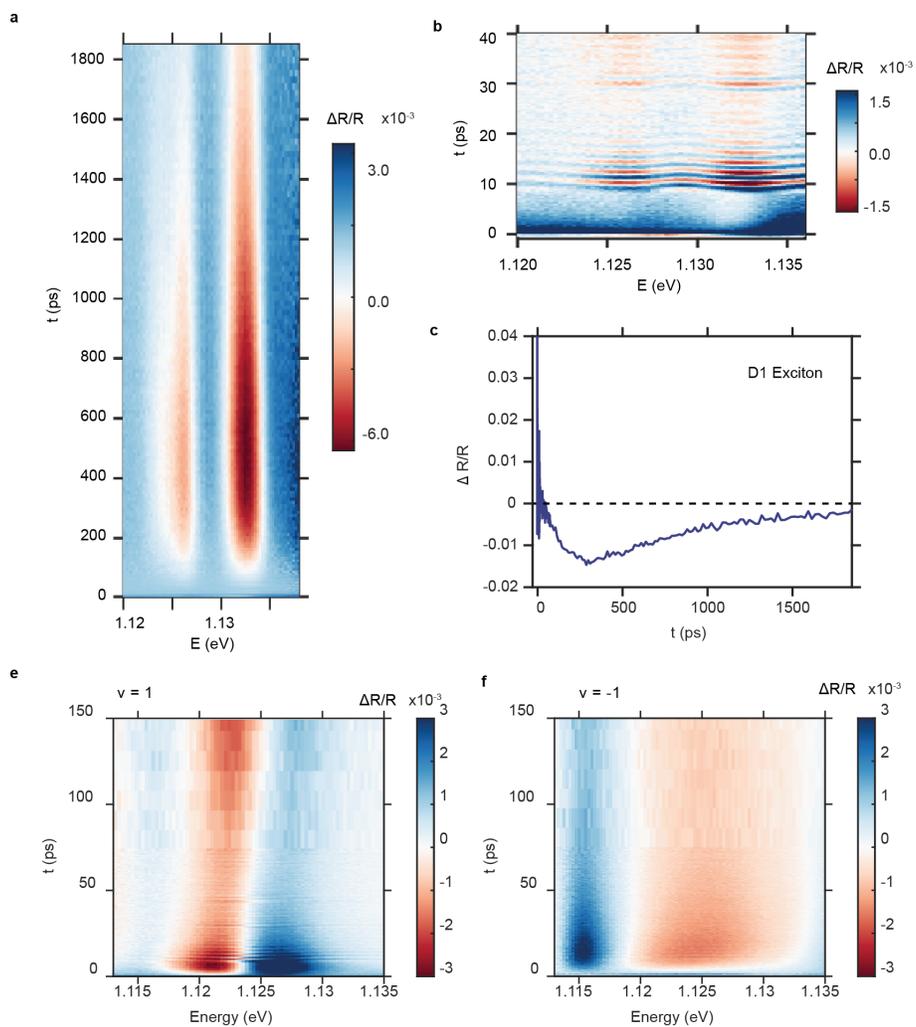

**Extended Data Fig. 11: spectral resolved time profile.** a. Transient reflection as a function of delay time and spectral energy for non-correlated state ν = 0 in D1. b. Short time window of a. c. Time profiles of the ν = 0 state at the exciton energy, only phonon modulation process observed. e. Transient reflection as a function of delay time and spectral energy for ν = 1 and f. ν = -1

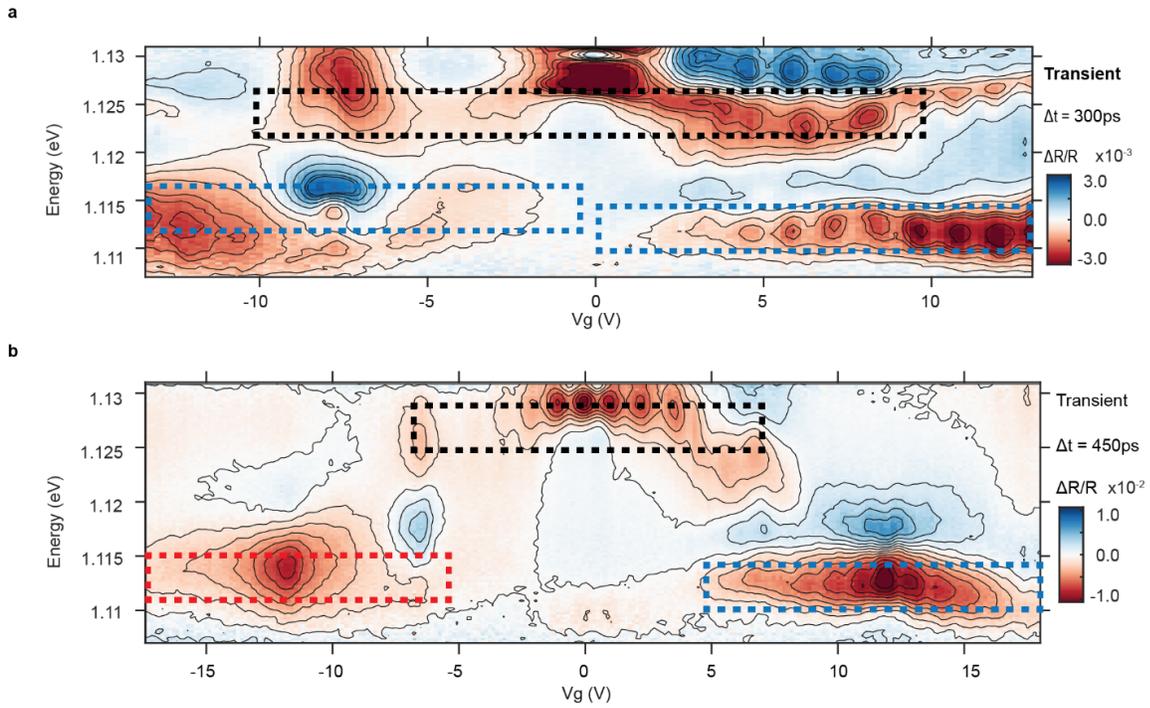

**Extended Data Fig. 12: Pump-probe spectroscopy at T = 2 K of tMoTe$_2$.** Transient reflection spectrum as a function of carrier density (n) and probe photon energy E ($\hbar\omega_2$) at pump-probe delays of **a.** Device D1 $\Delta t$ = 300 ps, with the dash line square indicating the integration range for the exciton (black) and trion (blue) shown in Figure 1g of the main text. **b.** Device D2 $\Delta t$ = 450 ps, with the dash line square indicating the integration range for the exciton (black), trion (blue), and trion derivative (red) shown in Figure 2d of the main text %% trion device 2 : 1.107—1.114 eV, exciton: 1.130 1.122, derivative: 1.115- 1.106 device 1: trion 1.108-1.114 exciton e1.120-1.125, exciton h 1.126-1.123, trion hole: 1.112-1.117

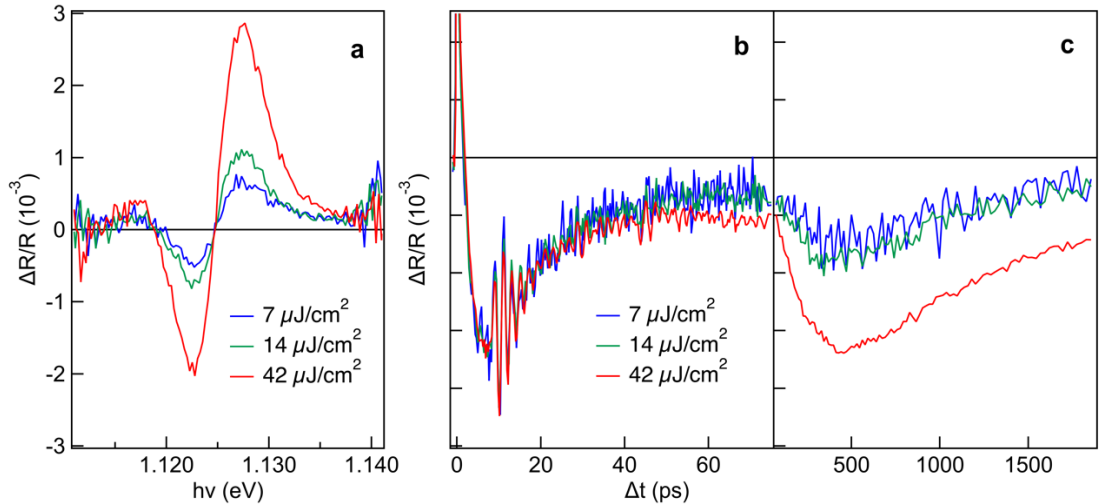

**Extended Data Fig. 13: Pump fluence dependent.** (a) Transient reflection in the exciton spectral region for the ν = 1 sate at pump fluences of ρ = 7, 14, and 42 μJ/cm$^2$ and a pump-probe delay time of $\Delta t$ = 7 ps. The magnitude of DR/R scales approximately linearly with ρ. Time profiles of the ν = 1 state at different fluences on two time scales (b) 0-75 ps and (c) 75-1875 ps. Each profile is obtained from integration of $\Delta$R/R in a probe photon energy window of 1.119-1.125 eV. All profiles in (b) and (c) are normalized to $\Delta$R/R at $\Delta t$ = 7 ps. The electronic melting/recovery processes are independent of r (b), while the relative magnitude of the phonon melting processes increases with pump fluence (c).

Extended table

Filling factor assignment:

| D1 | | | D2 | | |
|---|---|---|---|---|---|
| Filling factor (closest state) | Filling factor | Gate Voltage | Filling factor (closest state) | Filling factor | Gate Voltage |
| -5/3 | -1.779 | -13.600 | -8/3 | -2.702 | -16.295 |
| -3/2 | -1.589 | -12.150 | -5/2 | -2.500 | -15.095 |
| -4/3 | -1.350 | -10.325 | -7/3 | -2.239 | -13.540 |
| -1 | -1 | -7.644 | -2 | -1.967 | -11.918 |
| -2/3 | -0.663 | -5.069 | -5/3 | -1.636 | -9.940 |
| -3/5 | -0.568 | -4.344 | -3/2 | -1.493 | -9.090 |
| -1/2 | -0.501 | -3.834 | -4/3 | -1.327 | -8.100 |
| -2/5 | -0.402 | -3.079 | -1 | -1 | -6.511 |
| -1/3 | -0.316 | -2.418 | -4/5 | -0.802 | -5.242 |
| 2/7 | 0.280 | 2.243 | -2/3 | -0.664 | -4.336 |
| 1/3 | 0.345 | 2.756 | -1/2 | -0.497 | -3.235 |
| 2/5 | 0.404 | 3.234 | -1/3 | -0.334 | -2.157 |
| 3/7 | 0.469 | 3.758 | -1/7 | -0.163 | -1.032 |
| 1/2 | 0.478 | 3.820 | 0 | 0 | -0.028 |
| 4/7 | 0.578 | 4.622 | 1/7 | 0.155 | 1.063 |
| 2/3 | 0.729 | 5.836 | 1/3 | 0.327 | 2.198 |
| 3/4 | 0.732 | 5.856 | 4/7 | 0.541 | 3.608 |
| 6/7 | 0.879 | 7.033 | 2/3 | 0.661 | 4.402 |
| 1 | 1 | 8.000 | 6/7 | 0.829 | 5.510 |
| 6/5 | 1.190 | 9.522 | 1 | 1 | 6.672 |
| 4/3 | 1.331 | 10.645 | 6/5 | 1.163 | 7.709 |
| 3/2 | 1.493 | 11.943 | 4/3 | 1.333 | 8.830 |
| 5/3 | 1.625 | 13.000 | 3/2 | 1.506 | 9.970 |
| | | | 5/3 | 1.633 | 10.808 |
| | | | 9/5 | 1.764 | 11.675 |
| | | | 2 | 1.990 | 11.940 |
| | | | 11/5 | 2.182 | 13.070 |
| | | | 7/3 | 2.349 | 14.060 |
| | | | 5/2 | 2.494 | 14.910 |
| | | | 8/3 | 2.684 | 16.030 |